# Open and Linked Data Model for Carbon Footprint Scenarios

Boris Ruf, Marcin Detyniecki

*Research and Development, AXA Group Operations, Paris, France*

**Abstract**

Carbon footprint quantification is key to well-informed decision making over carbon reduction potential, both for individuals and for companies. Many carbon footprint case studies for products and services have been circulated recently. Due to the complex relationships within each scenario, however, the underlying assumptions often are difficult to understand. Also, re-using and adapting a scenario to local or individual circumstances is not a straightforward task. To overcome these challenges, we propose an open and linked data model for carbon footprint scenarios which improves data quality and transparency by design. We demonstrate the implementation of our idea with a web-based data interpreter prototype.

## 1. Introduction

Recent accumulation of natural disasters has accelerated awareness of the unfolding climate crisis. The extreme weather events have also increased motivation in society to take action [1, 2]. While the important role of cutting greenhouse gases (GHGs) in mitigating the effects of global climate change is well understood, identifying the individual reduction potential is less obvious: The subject is complex, involving hidden, embedded emissions that often occur somewhere else and are hard to visualize. Robust, illustrative comparisons with common baseline activities could help put the magnitude of a specific emission scenario in perspective and support the formulation of clear and plausible recommendations. In the last few years, catchy claims have received a lot of attention, but they have ultimately proven to be flawed. One such statement went that a typical year of incoming e-mail accounted for 136 kg of emissions, or the equivalent of "driving 200 miles in an average car" [3]. The author has dissociated himself from the analysis which was not rigorous and subject to accumulated rounding-off errors [4]. Another popular statement went that watching 30 minutes of Netflix was equivalent to "driving almost 4 miles". This claim was derived from a study [5] which was later found to contain a far-reaching bit/byte conversion error on the one hand, but also missed a major source of emission, namely the end user device [6]. These incidents underline the need for transparent methodologies to assess and discuss carbon footprint scenarios. The default approach is to use life cycle assessment (LCA) [7], a instrument of choice in many current carbon footprint studies (see [8, 9, 10] for examples). It is a standardized and solid but also complex methodology which is difficult to decompose and adapt. Simulation models have been presented as alternative approach, however, limited to transport and store functions [11]. In the following, we propose a new, simple data model which is tailor-made for carbon footprint scenarios and has a few domain-specific features.

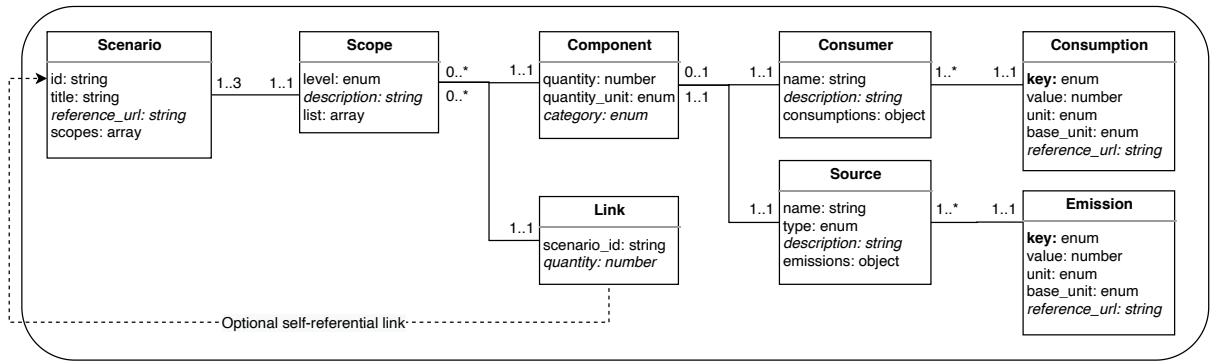

Fig. 1. Data model of self-referential carbon footprint scenario (optional attributes are italic).

## 2. Data model

To improve accessibility and utility of carbon footprint data, we pursue an open data approach, as envisioned by the EU open data strategy [12]. To facilitate sharing and re-usability, we adopt the concept of "linked data" [13] which was designed to interlink machine-readable data on the Internet. Such a modular architecture allows for distributed data hosting. It also enables the presentation of varying levels of detail which can be navigating through the nested user interface. As data format, we chose JSON [14] for its lightweight and compact style.

The schema in Fig. 1 outlines our proposal of a universal carbon footprint scenario data model. The boxes represent the entities which include a list of attributes, followed by their data types. Optional attributes are shown in italics. The connectors specify the relationships between the entities. Numbers on the connectors express the minimum and maximum cardinalities of the relationship.

Any **Scenario** is identified by a Uniform Resource Identifier (URI). It features a title which optionally includes a reference in form of a hyperlink. A scenario covers 1–3 **Scopes** of emission as defined by the GHG Protocol [15]. This entity may contain a description. Further, it includes 1 or several **Components** or **Links**. The latter simply relates to another scenario which is identified by its URI. It may be supplemented by an indication of quantity. A component, however, must include a quantity and a quantity unit (e.g., "km", "kg", "pcs"). It can also have a category which defines the type of consumer (e.g., "car", "food", "electronics"). Further, this entity must include a **Source** which has a name and a type (e.g., "France electricity grid" and "electricity", or "premium gasoline" and "gasoline"). It can also include a description (e.g., "Year 2022"). Any source must include 1 or more **Emissions**, implemented as key-value pairs. The type of emission defines the key ("co2e", "co2", "ch4", "n2o", "hfcs", "pfcs", "sf6", "nf33"). Value, unit (e.g., "g", "kg", "t") and base unit (e.g., "kWh", "l", "kg", "km") specify the emission details. A reference to the source of this information can be disclosed as hyperlink. The component may also have a **Consumer** when information about the energy efficiency is available (consider that emissions of food, for example, often are reported only per 1 kg produced). A consumer has a name (e.g., "Boeing 747", "iPhone 14"), and optionally a description. It also includes 1 or more **Consumptions** (some consumers may support several energy sources, for example cars with combustion engine can be refueled with different types of gasoline; hybrid vehicles use gasoline and electricity) as key-value pairs. The type of energy defines the key (e.g., "electricity", "gasoline") and corresponds to the type of the source. Value, unit (e.g., "kWh", "l") and base unit (e.g., "km", "h", "d") specify the consumption details. This information can be supplemented by a reference.

For a full code sample please see Listing 1 in the appendix of this document.

## 3. Prototype

We have implemented a web-based prototype to evaluate the feasibility of our proposed data model. It works as interpreter for the data to view and explore the carbon footprint scenario. It also adds a couple of features for the user to interact with the data and re-use it. We will explain them in detail in the corresponding section below. A screenshot of the application can be found in Fig. 2 a).

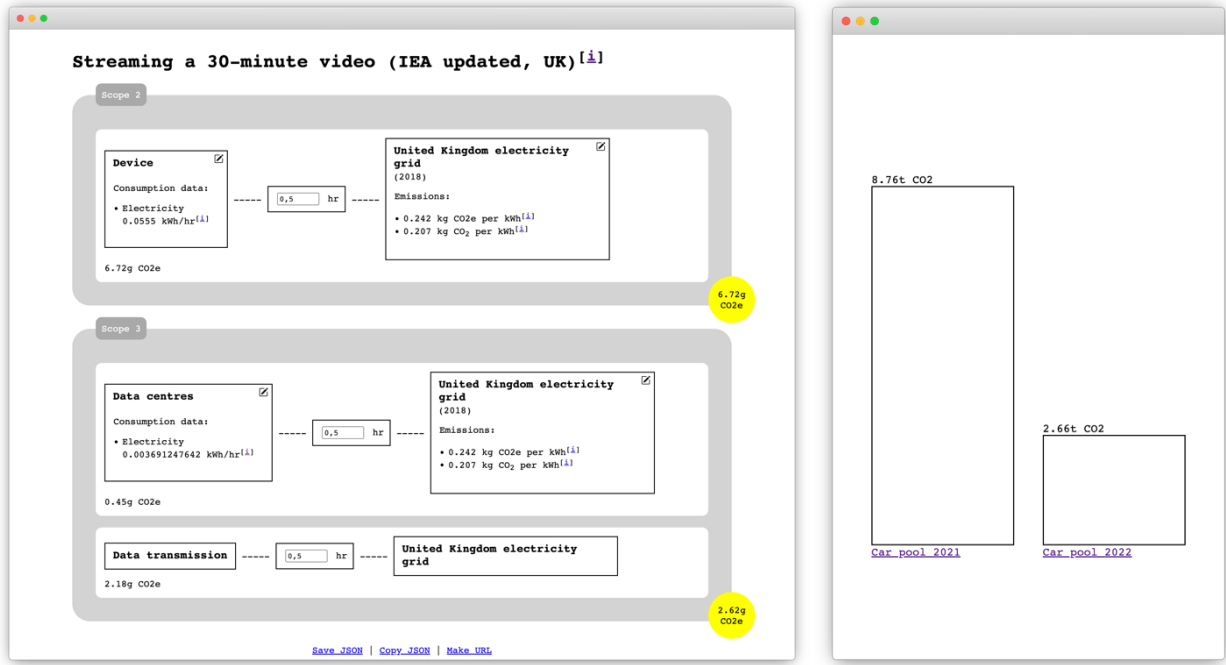

Fig. 2. Screenshots of web-based data interpreter in (a) viewer mode and in (b) benchmark mode.

### 3.1. Technical notes

The prototype was implemented in Javascript, running serverless as GitHub page. This sets a very low bar for deployment and advancement of the viewer. Due to the self-referential structure of the data model, the nested scenarios can be hosted in a distributed manner. The data interpreter fetches and processes them recursively. A more detailed illustration of an exemplary transaction flow can be found in Fig. 3 in the appendix.

### 3.2. Features

The main purpose of the viewer application is to interpret and render the data of a carbon footprint scenario provided in the format of the data model introduced in section 2. The viewer aggregates the emission data per element and per scope, while handling unit conversion and finding a common ground based on the availability of emission data by type of emission (e.g., "CO2e", "CO2"). Further, the user can manipulate the data in the user interface with immediate impact on the results: For each element, quantities can be adjusted. It is also possible to connect different data sources to replace consumer components and energy sources on-the-fly. Data of scenarios that were customized in the viewer can get downloaded as JSON file. They can also get encoded as URL and shared, which makes collaboration on a scenario very convenient. Lastly, a benchmark view, as shown in Fig. 2b), allows for easy comparison of 2 or more scenarios by means of their identifiers.

*3.3. Examples*

Several sample scenarios are available for testing in our GitHub repository[1]. This is also where to find more detailed documentation on how to use and deploy the prototype. The screenshot in Fig. 2a) shows a scenario modelling the estimated carbon footprint linked to 30 minutes of online video streaming[2]. It was adopted from a study by the International Energy Agency (IEA) [6]. Emissions caused by the end user device are in scope 2, data centers and data transmission account for scope 3 emissions. Consumption and emission factors are referenced accordingly. The user can interact with the scenario and test different configurations by modifying the end user device, the duration, and the electricity mix. Customized scenarios can get exported and shared using the links on the bottom of the page.

## 4. Conclusion

We present an open and linked data scheme to model carbon footprint scenarios. To evaluate our proposal, we developed a web-based data interpreter prototype. Several features introduce positive changes to the status quo. The possibility to provide references to the source of information for all underlying assumptions gives transparency and traceability a boost. The nested, self-referential approach allows to explore a carbon footprint scenario on different levels of detail. It also provides an increased degree of re-usability and adaptability. Finally, automated unit conversion and common emission type detection help avoid data conversion errors. With this work, we hope to contribute to the development of a standard data model for carbon footprint quantification.

**Appendix A.    Code sample**

```json
{
  "title": "Mobility",
  "scopes": [
    {
      "level": "Scope 1",
      "list": [
        {
          "type": "component",
          "consumer": {
            "name": "Volkswagen Golf (2014)",
            "description": "Engine ID 45, 4 cylinders, Manual 6-spd",
            "consumptions": {
              "diesel": {
                "value": "0.0735046875",
                "unit": "l",
                "base unit": "km",
                "reference_url": "https://www.fueleconomy.gov/"
              }
            }
          },
          "quantity": "10000",
          "quantity_unit": "km",
          "source": {
            "name": "Gas/Diesel oil",
            "type": "diesel",
            "emissions": {
              "co2e": {
                "value": "3.25",
                "unit": "kg",
                "base unit": "l",
                "reference_url": "https://bilansges.ademe.fr/index.htm?new_liquides.htm"
              }
            }
          }
        }
      ]
    }
  ]
}
```

*Listing 1. Code sample (JSON)*

---

1 https://github.com/borisruf/carbon-footprint-modeling-tool
2 https://borisruf.github.io/carbon-footprint-modeling-tool/?id=scenario-8f35af7c-ee5b-42aa-b538-371b126b3d24

**Appendix B.    Transaction flow chart**

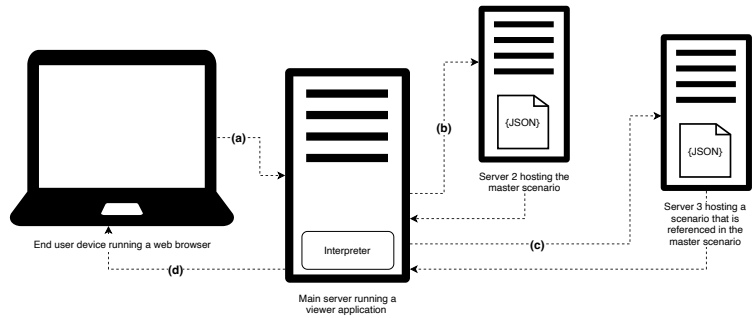

*Fig. 3 Exemplary illustration of transaction flow running to load and compute distributed carbon footprint scenario:*
*a) A user accesses the website of a viewer application providing the URI of the master scenario in the URL.*
*b) The main server locates the referenced scenario and loads the corresponding JSON file from another server.*
*c) After processing the data, the main server loads another scenario which is linked to in the master scenario.*
*d) The main server recursively calculates all emission values and renders the full scenario in the web browser.*